\documentclass[showpacs,twocolumn,amssymb,prl,aps]{revtex4}
\usepackage{graphicx,epsfig}
\usepackage{upgreek}
\usepackage{wasysym}

\newcommand{\dC}{\,$^{\circ}$C}

\newcommand{\etal}{\textit{et al.}}

\begin{document}
\title{Spontaneous Imbibition Dynamics of an n-Alkane in Nanopores:\\ Evidence of Meniscus Freezing and Monolayer Sticking}
\author{Simon Gruener}
\author{Patrick Huber}
\affiliation{Faculty of Physics and Mechatronics Engineering, Saarland University, D-66041 Saarbr\"ucken (Germany)}

\date{\today}

\begin{abstract}
Capillary filling dynamics of liquid n-tetracosane (n-C$_{\rm 24}$H$_{\rm 50}$) in a network of cylindrical pores with 7 and 10~nm mean diameter in monolithic silica glass (Vycor) exhibit an abrupt temperature-slope change at $T_{\rm s}=54\,^{\circ}$C, $\sim 4\,^{\circ}$C above bulk and $\sim 16\,^{\circ}$C, $8\,^{\circ}$C, resp., above pore freezing. It can be traced to a sudden inversion of the surface tension's $T$-slope, and thus to a decrease in surface entropy at the advancing pore menisci, characteristic of the formation of a single solid monolayer of rectified molecules, known as surface freezing from macroscopic, quiescent tetracosane melts. The imbibition speeds, that are the squared prefactors of the observed square-root-of-time Lucas-Washburn invasion kinetics, indicate a conserved bulk fluidity and capillarity of the nanopore-confined liquid, if we assume a flat lying, sticky hydrocarbon backbone monolayer at the silica walls.
\end{abstract}



\maketitle

When a liquid wets the inner wall of a capillary, it forms a concave meniscus and spontaneously invades, a phenomenon called capillary filling or spontaneous imbibition (SI). For channels with diameters of a few nanometers the interfacial forces driving this process can defy gravitational and viscous forces by orders of magnitude \cite{Nanofluidics}. Therefore, the study of SI in such extremely spatially confined geometries has attracted increasing interest, both from experiment \cite{CRExp} and theory \cite{Dimitrov2007, CRTheo}. Typical curvature radii of the advancing menisci are expected to be on the order of the pore radius \cite{Dimitrov2007} and thus on the order of the size of the building blocks of the invading liquids. This may alter the equilibrium structural properties of the liquids adjacent to the menisci and the flow characteristics in their proximity. For example, recent molecular dynamics studies show peculiar molecular rearrangements at advancing imbibition fronts of models for polymer melts \cite{Dimitrov2007}. Moreover, for a pore of a few nm in diameter any change in the no-slip velocity boundary condition, by an in- or decreased mobility of the molecules in the wall proximity, is expected to alter SI dynamics markedly \cite{Dimitrov2007, CRTheo}.

\begin{figure}[htbp]
\includegraphics[width=.65\columnwidth]{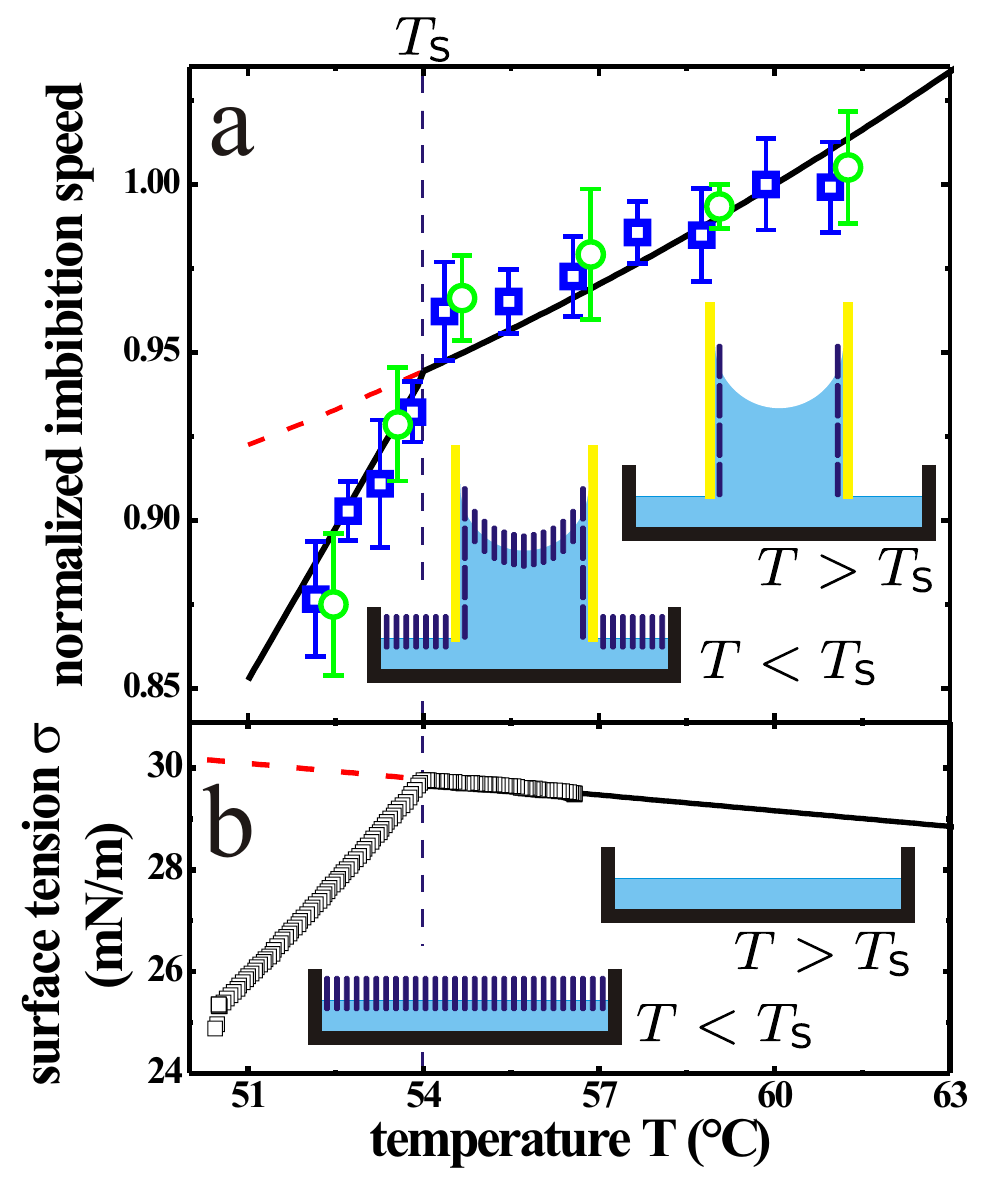}
\caption{(color online). (a) Measured normalized imbibition speeds $v_{\rm i}^{\rm n}$ (definition see text) of C24 in Vycor with 3.4~nm ($\square$) and 4.9~nm ($\fullmoon$) mean pore radius, resp., in comparison with values calculated under the assumption of meniscus freezing (--). Insets: Illustration of a SF-layer at an advancing meniscus ($T<T_{\rm s}$) and of a sticky, flat lying boundary layer at the silica pore wall in the entire $T$-range investigated. (b) $T$-dependent surface tension of C24 (courtesy of M. Deutsch). Insets: Illustrations of SF at a planar C24 surface \label{fig2}. The dashed lines correspond to extrapolations of the calculated imbibition speeds and of the surface tension below $T_{\rm s}$ in the absence of surface freezing, respectively.}
\end{figure}

Here we present a $T$-dependent experimental study of SI of the linear hydrocarbon n-tetracosane (n-C$_{\rm 24}$H$_{\rm 50}$, denoted C24 in the following) in silica nanopores. C24 exhibits surface freezing (SF) \cite{SFnAlkane}, that is the formation of a single solid monolayer floating on the bulk melt in a $T$-range between the bulk freezing temperature $T_{\rm f}$ and a temperature $T_{\rm s}$. Upon SF the C24 molecules are rectified, parallel-aligned with their long axis along the surface normal and the center of mass lattice is hexagonal, resulting in a 3~nm thick surfactant-like monolayer (see illustration in Fig.~\ref{fig2}). The abrupt onset of SF at $T_{\rm s}$ allows one to switch on (and off) the 2D crystalline phase by a small $T$-variation. 
The goal of our study is to investigate whether this peculiar molecular rearrangement is detectable by, and how it affects the imbibition dynamics of, C24 in nanopores. Some attention will also be paid to possible velocity slippage at the pore walls, and thus to deviations from macroscopic SI theory.

\begin{figure}[htbp]
\includegraphics[width=0.7\columnwidth]{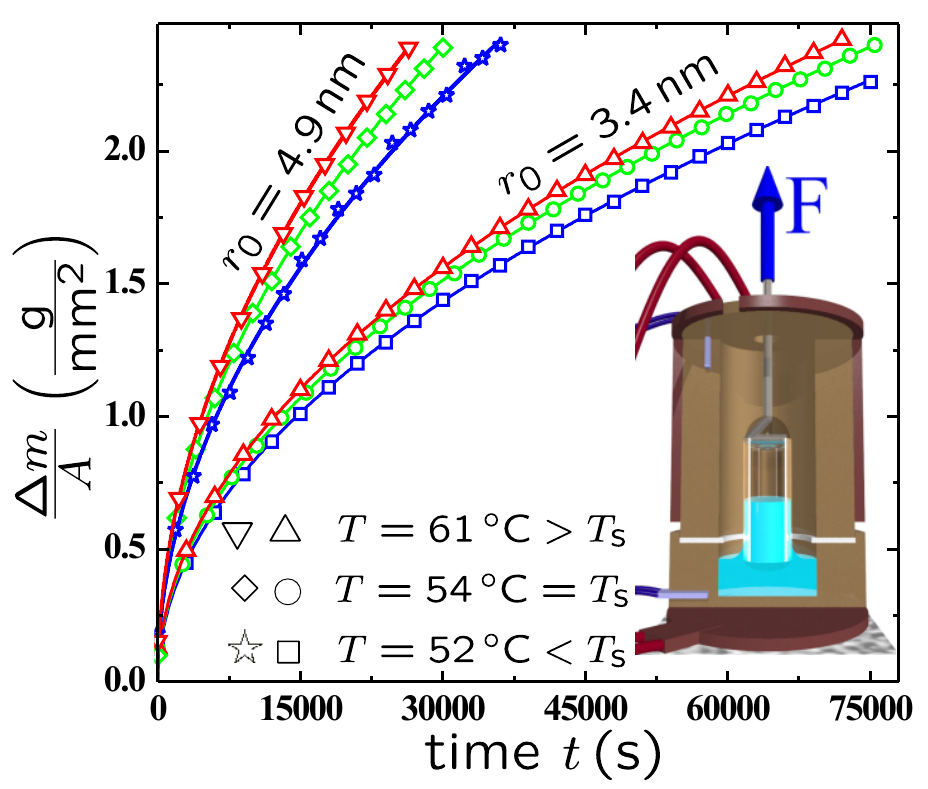}
\caption{\label{fig1} (color online). Specific mass uptake of Vycor for two mean pore radii $r_{\rm 0}$ due to C24 imbibition as a function of time for selected $T$'s close to, but above the bulk freezing temperature. Solid lines correspond to $\sqrt{t}$-fits discussed in the text. Inset: Raytracing illustration of the thermostated imbibition cell employed for the isothermal capillary rise experiments.}
\end{figure}

We selected C24 for our study, since it exhibits the largest SF $T$-range of all pure n-alkanes ($T_{\rm s} = 54$\dC\ and $T_{\rm f} = 50$\dC) \cite{SFnAlkane}. As a mesoporous host we have chosen Vycor glass (Corning glass) from two sample batches differing in the mean pore diameter, only. It is a virtually pure fused silica glass permeated by a 3D network of cylindrical, tortuous pores \cite{Levitz1991}. We cut blocks of $\sim 10$\,mm height from the cylindrical rods (cross sectional area $A\approx 37\,{\rm mm}^2$). After cleaning \cite{Gruener2009} the samples were attached to an electronic balance with a resolution of $10^{-2}$~mg and were kept in a closed, $T$-controlled copper cell with a $T$-stability of 0.1\dC, see inset of Fig.~\ref{fig1}. N$_{\rm 2}$ sorption experiments were performed at $-196$\dC\ and analyzed within a mean-field model for capillary condensation \cite{SC} yielding a mean pore radius of $r_{\rm 0}= 3.4~{\rm nm}$ and $4.9~{\rm nm}$ for samples from the two Vycor batches, resp., and an identical volume porosity $\phi_{\rm 0}= 0.31$. 

The mass uptake of Vycor as a function of imbibition time, $m(t)$, was recorded gravimetrically after bringing the Vycor sample in contact with the liquid surface. It is plotted, normalized by $A$, as a function of time $t$ for a selected set of $T$'s in Fig.~\ref{fig1}. We find monotonically increasing $m(t)$-curves and the mass uptake rate decreases with $t$. Neglecting the complex pore structure, one can understand the imbibition dynamics by simple phenomenological considerations: If a liquid wets or partially wets a capillary with radius $r$ (contact angle with the pore wall, $\Theta < 90 ^\circ$), it is sucked by the capillary pressure, $p_{\rm c}$, given by the Laplace equation, $p_{\rm c} = 2 \,\sigma \,\cos (\Theta)/r$. The tube geometry calls for a Hagen-Poiseuille type flow pattern, where the viscous drag implies that the volume flow rate is $\dot{V}(t) \propto p_{\rm c}/({\eta}\, l(t))$ and thus the corresponding mass flow rate is given by $\dot{m} \propto \rho\, p_{\rm c}/(\eta \, l(t))$, where $\rho$ is the mass density of the liquid and $l(t)$ refers to the height of the column that has already been filled up with liquid at time $t$. This length is proportional to $m(t)$. Hence a quite simple relation for $m(t)$ and its time derivative is arrived at: $m(t)\,\dot{m}(t) \propto \rho^2\, p_{\rm c}/\eta  \propto \rho^2 \,\sigma\,\cos(\Theta)/\eta  \label{Gl1}$. It is solved by
\begin{eqnarray}
m(t)  \propto \rho  \underbrace{\sqrt{ \frac{ \sigma}{\eta} \cos(\Theta)}}_{\sqrt{v_{\rm i}(T)}} \;\; \sqrt{t} \; . \label{Gl2}
\end{eqnarray}
In a macroscopic SI experiment gravitational forces are to be considered, as well. However, for the tiny tubes or pores investigated here, the hydrostatic pressure is negligible compared to $p_{\rm c}$. Consequently the mass uptake in a SI experiment should show a $\sqrt{t}$-behavior and it is sensitive to the fluid parameters [$\sigma(T)$, $\eta(T)$, $\rho(T)$] and the fluid wall interaction (via the contact angle $\Theta$). Relation (\ref{Gl2}) is known as Lucas-Washburn behavior \cite{Lucas1918} and the square of the second prefactor as imbibition speed, $v_{\rm i}(T)$. Porous media can be considered as complex networks of simple capillaries, which adds an additional $t$-independent geometry factor to relation (\ref{Gl2}). Therefore, we expect a $\sqrt{t}$-behavior for $m(t)$, too. In accordance with this statement, the data points of our experiments show excellent agreement with $\sqrt{t}$-fits for the entire SI time and temperature range investigated (see Fig.~\ref{fig1}).\\
Since we used matrices with identical pore microstructures for all isothermal SI experiments we can eliminate the unknown geometry factor of relation (\ref{Gl2}) by a calculation of normalized isothermal imbibition speeds. We divide the prefactors necessary to fit the measured $m(t)$ curves by the prefactor value at an arbitrarily chosen $T$ (here $T^{\rm n}=60$\dC) and arrive at the normalized values $v^{\rm n}_{\rm{i}}(T)$, plotted in Fig.~\ref{fig2}(a) in the vicinity of $T_{\rm f}$. They are monotonically decreasing with decreasing $T$, which is attributable to the monotonical increase in $\eta(T)$ upon cooling \cite{Beiner2004}. Note, however, that at $T=T_{\rm s}=54$\dC\ we observe an abrupt change in the $T$-slope of $v^{\rm n}_{\rm{i}}(T)$.\\
In order to analyze this peculiarity we recall the distinct signature of SF in the $T$-behavior of $\sigma$, known from bulk n-alkanes, that is a change from a small negative, above $T_{\rm s}$, to a large positive $T$-slope, below $T_{\rm s}$ [see Fig.~\ref{fig2}(b)]. It can be understood from elementary surface thermodynamics \cite{SFnAlkane}: The $T$-slope of $\sigma(T)$ yields information on the surface excess entropy: ${\rm d}\sigma/{\rm d}T=-(S_{\rm s}-S_{\rm b})$, where $S_{\rm s}$ and $S_{\rm b}$ are the entropies of the surface and bulk, respectively. The negative slope of bulk C24 for $T>T_{\rm s}$ is typical of an ordinary liquid surface where the molecules on the surfaces are less constrained than those in the bulk, thus $S_{\rm s}$ is slightly larger than $S_{\rm b}$, yielding ${\rm d}\sigma/{\rm d}T<0$. SF and its first-order character results in an abrupt reduction of the surface entropy $S_{\rm s}$ in such a way that $S_{\rm s}$ is smaller than $S_{\rm b}$ leading to ${\rm d}\sigma/{\rm d}T>0$.
Assuming this $\sigma(T)$ kink-behavior along with the $T$-dependency of $\eta$ as measured with a cone-plate rheometer \cite{Beiner2004}, we calculated normalized imbibition speeds $v_{\rm i}^{\rm theo}(T)$, plotted as a solid line in Fig.~\ref{fig2}(a). The quantitative agreement with our measured changes in the imbibition dynamics is reasonable. Note that a change in the n-alkane/silica interaction at $T_{\rm s}$ or pore clogging may also be responsible for the behavior observed. However, $T$-dependent contact angle measurements on droplets of C24 on planar silica, sensitively depending on this interaction, showed $\Theta(T)$ of 6$\pm 2^{\circ}$ ($\cos \Theta \sim 1$) and no measurable change upon crossing $T_{\rm s}$. Clogging and the corresponding decrease in $v_{\rm i}$ should be more pronounced in the smaller pores, which we also did not observe. Therefore, we feel encouraged to solely attribute the distinct change in the SI dynamics to a change in $\sigma(T)$ at the advancing menisci characteristic of SF. Moreover, the slope difference in $\sigma(T)$ inferred from linear fits of our measured imbibition speeds below and above $T_{\rm s}$ of $1.75\pm 0.3$~mN/(m$\cdot$K) is within error margins equal to $1.40 \pm 0.1$~mN/(m$\cdot$K) for SF at bulk C24 melts. Since this quantity measures the loss of entropy of surface molecules upon entering the SF state: $\Delta ({\rm d} \sigma/{\rm d}T)={\rm d} \sigma/{\rm d}T (T<T_{\rm s})-{\rm d} \sigma /{\rm d}T(T>T_{\rm s})=\Delta S$, our measurements indicate that the surface ordering at the advancing imbibition front leads to a loss in entropy comparable to the one observed at quiescent bulk surfaces and thus, presumably, to a well ordered 2D crystalline state. Having in mind the dynamics of the frozen meniscus along the tortuous pore path, the frequent encountering of pore junctions with irregular channel geometries and the small curvature radii of the menisci, where large defect densities of the 2D crystals are expected \cite{Bowick2000}, this may seem surprising. Note, however, the defect formation rate is presumably proportional to the monotonically in $t$ decreasing meniscus velocity, $ \sim {0.21}/{\sqrt{t\,[{\rm ns}]}}$\,1/ns expressed here in terms of C24 all-trans length of 3~nm. This rate falls below typical crystallization speeds of n-alkanes of 0.1 C24 length/ns \cite{Turnbull1961} already after a few nanoseconds of elapsed SI time. Therefore, it is plausibel that the fast crystallization kinetics can heal out any occurring defects on $t$-scales much smaller than detectable by SI dynamics.\\
SF at bulk n-alkane surfaces occurs over a small $T$-range $\left| T_{\rm f}-T_{\rm s} \right|$ of a few degrees celsius \cite{SFnAlkane}. However, $T_{\rm m}$ is expected to be significantly shifted downwards in pore-confinement \cite{AlbaSimionesco2006}. X-ray diffraction experiments on C24 in 3.5~nm and 4.5~nm Vycor (see \cite{Huber2004}) indicated freezing of the pore-confined liquid at $T=38$\dC and~$42$\dC, only. Thus, our evidence of an unchanged $T_{\rm s}$ despite the sizeable downward-shift of pore freezing suggest that pore-confinement may allow one to establish SF over a much larger $T$-range than known of any other free surface of bulk n-alkane melts, i.e. $T_{\rm s}-T_{\rm m}$ $\sim 16$\dC. Unfortunately, we did not find unambigious hints of the Bragg peak pattern typical of the SF state in those diffraction experiments. The dominating 70\% volume fraction of silica along with the coincidence of the first maximum of the structure factor of both the silica and the nano-confined liquid with the position of the dominating SF-Bragg peak render its detection in Vycor particularly challenging. However, in agreement with our conjecture on the extended range of the SF state, this signature has been observed in a $T$-range from $T_{\rm s}$ down to the pore freezing temperature in diffraction studies on silicon nanochannels partially filled with n-alkanes (10~nm diameter and thin silica pore wall corona) \cite{Henschel2007}. 

As outlined in Ref.~\cite{Gruener2009}, more detailed insights regarding an SI experiment can be achieved by extending Darcy's law for the volume flow rate $\dot V$ through a surface area $A$ of a porous host. One arrives at the following generalized equation for the Lucas-Washburn law:
\begin{equation}
\label{Eqmt}
m^2(t)= \rho^2\,A^2\,\frac{r_{\rm h}^4\,\phi_0^2}{2\, r_0^3 \,\tau} \frac{\sigma}{\eta} t.
\end{equation}

The tortuosity $\tau$ characterizes the connectivity and meandering of the pores. It could be inferred from experiments on the self-diffusion of liquids in porous Vycor \cite{Lin} to $\tau=3.6 \pm 0.4$, in accordance with simulations of its pore morphology \cite{Crossley}. Furthermore, in Eq.~(\ref{Eqmt}) $r_{\rm h}$ refers to the hydrodynamic radius over which a parabolic flow profile is expected in the pores. It does not necessarily have to agree with the mean pore radius $r_{\rm 0}$ because of either strongly adsorbed, immobile boundary layers, or due to velocity slippage at the pore walls \cite{slip, Dimitrov2007}. Using the parameters of the pore morphology determined from our sorption isotherm measurements ($r_{\rm 0}= 3.4~{\rm nm}$, $\phi_{\rm 0} = 0.31$) we overestimate the absolute imbibition speed compared to the measured ones for all $T$'s by $\sim 30\,\%$, well beyond our error margin of $\sim 10\,\%$ in the SI experiment. However, if we assume a hydrodynamic radius $r_{\rm h}$ which is reduced by the thickness of approximately 1 hydrocarbon backbone, that is 0.4~nm, along with a $\sigma(T)$ change characteristic of SF, we arrive at a good agreement between measured and predicted $m(t)$ curves for all $T$'s investigated (see Fig.~\ref{fig1}). More importantly, also for SI in a sample of the second Vycor batch with $r_{\rm 0}=4.9$ nm, the assumption of an identically reduced effective hydraulic diameter $r_{\rm h}=r_{\rm 0}-0.4$~nm$=4.5~$nm can account for the measured SI dynamics, see Fig.~\ref{fig1}. The assumption of a flat lying, sticky monolayer is corroborated by forced imbibition experiments on n-alkanes in Vycor \cite{Debye1959} and by studies regarding the thinning of n-alkane films in the surface force apparatus \cite{SFA}. Moreover, X-ray reflectivity studies indicate one strongly adsorbed, flat lying monolayer of hydrocarbons on silica \cite{nAlkaneSilicaStructure}. The evidence of SF in nanopores, presented above, are also supported by a study on capillary condensation of hydrocarbons in slit geometry \cite{Maeda2000}. Also in semi-confined, planar, thin film geometries on silica surfaces indications of SF have been discussed with regard to the complex wetting and freezing behavior of n-alkanes \cite{SFThinFilm, SFWettingFreezing}.\\
We reported a first evidence of SF occurring at advancing menisci in silica nanopores. This observation testifies a remarkable robustness of this archetypical surface ordering transition both upon nanoscale spatial confinement and upon self-propelled movement of the interface in a rather complex, tortuous pore network. The influence of SF on the SI flow dynamics can entirely be accounted for by the change in surface tension typical of SF and does not additionally slow-down the meniscus movement. This finding supports an impressive flexibility of the SF layer similarly as it has been inferred from capillary wave spectroscopy \cite{Hughes1993}. The SF layer can be considered as a peculiar case of a surfactant. Therefore, our study examplifies also how the formation of a surfactant layer at nanoscopic menisci can markedly affect the imbibition dynamics in nanocapillaries, similarly as it has been explored for macropores \cite{LabajosBroncano2006}. For the future, we envision experiments on the huge variety of systems exhibiting SF, which ranges from alcohols, semi-fluorinated alkanes, and diols to polymers comprising alkyl chains in the backbone or as side chains to liquid alloys, liquid crystals and tetrahedral liquids \cite{OSSF}. The SI dynamics of all these technologically important systems may be affected by SF. Our study also suggests that nanopore confinement may allow one to establish the SF state over much larger $T$-ranges than possible at bulk surfaces. Therefore, our findings may motivate studies how SF influences pore crystallization, given the proven impact of SF on n-alkane crystallization both in the bulk and in the spatially confined state \cite{Kraack2000}. Finally, we believe molecular dynamics simulations in combination with synchrotron x-ray diffraction experiments and optical birefringence measurements on aligned tubular silica channels \cite{Kityk2008} may allow one to elucidate how the SI velocity profiles at the advancing menisci will be altered by a SF layer, how the SF layer meets the confining silica walls and how the available theories and simulations of SF \cite{SFSim} can be extended towards non-equilibrium conditions typical of SI investigated here. 

\indent We acknowledge financial support by the DFG (priority program 1164, \textit{Nano- and Microfluidics}).

\end{document}